\documentclass{article}
\usepackage{amsfonts}
\usepackage{amssymb}
\usepackage{caption}
\usepackage[dvips]{graphicx}

\begin{document}

\title{Lie-Algebraic Interpretation of the Maximal
Superintegrability and Exact Solvability of the Coulomb-Rosochatius
Potential in $n$ Dimensions}

\author{G. A. Kerimov$^{1}$ and A. Ventura$^{2,3}$\\
${}^1$ Physics Department, Trakya University, Edirne, Turkey\\
${}^2$ ENEA, Centro Ricerche Ezio Clementel, Bologna, Italy \\
${}^3$ Istituto Nazionale di Fisica Nucleare, Sezione di Bologna,Italy}

%\date{\nodate}

\maketitle

\begin{abstract}
The potential group method is applied to the $n$-dimensional
Coulomb-Rosochatius potential, whose bound states and scattering states are
worked out in detail. As far as scattering is concerned, the $S$-matrix
elements are computed by the method of intertwining operators and an
integral representation is obtained for the scattering amplitude. It is
shown that the maximal superintegrability of the system is due to the
underlying potential group and that the $2n-1$\ integral of motions are
related to Casimir operators of subgroups.
\end{abstract}

%\pacs{03.65.Fd, 03.65.Ge, 03.65.Nk}
%\maketitle

%\address{$^{1}$Physics Department, Trakya University, Edirne, Turkey} %
%\address{$^{2}$ENEA, Centro Ricerche Ezio Clementel, Bologna, Italy} 
%\address{$^{3}$Istituto Nazionale di Fisica Nucleare, Sezione di Bologna,
%Italy}

%\submitto{\JPA}

\section{Introduction}

The Coulomb system in arbitrary dimensions is among the well-known and best
studied exactly solvable systems in quantum mechanics\cite{Al58}-\cite{No99}%
. It describes the dynamics of charged particles under the influence of the $%
1/r$ \ potential. This is what we conventionally mean by Coulomb system,
even if the true $n$-dimensional Coulomb potential satisfying the Poisson
equation goes like $1/r^{n-2}$, for $n\geq 3$.

It has been shown that the bound and scattering states of a Coulomb system
in $n$\ dimensions can be associated with the rotation group $SO\left(
n+1\right) $\ and the Lorentz group $SO\left( n,1\right) ,$\ respectively%
\cite{Al58}-\cite{RS79}. In other words, for a fixed energy, these groups
appear as invariance groups of the system. These symmetries for $n=3$\ were
discovered by Fock\cite{Fo35} and Bargmann\cite{Ba36}. Moreover, it was
pointed out in Ref.\cite{Th89} that the dynamical group that yields the
positive and negative energy spectra of the Coulomb problem in $n$\
dimensions is $SO\left( 2,n+1\right) $,\ \ while the dynamical group in the
case $n=3$\ had been discussed by Barut\cite{BA72}. It is also worth noting
that Coulomb and harmonic-oscillator systems are the only spherically
symmetric superintegrable systems in arbitrary dimensions.

We recall that in classical mechanics a closed system with $n$\ degrees of
freedom is completely integrable if it admits $n$\ integrals of motion
(including the Hamiltonian ) that are independent and in involution , i.e.
the Poisson brackets of any two integrals are zero. The system is called
superintegrable if there exist $q$, $1\leq q\leq n-1$, additional
independent integrals of motion. The cases $q=1$\ and $q=n-1$\ correspond to
minimal and maximal superintegrability, respectively. In quantum mechanics
the definitions of complete integrability and superintegrability are same,
but Poisson brackets are replaced by commutators (see Ref.\cite{Te04} for a
general review).

In the present work we shall consider a quantum system characterized by a
Hamiltonian of the form 
\begin{equation}
H=\frac{1}{2}p^{2}-\frac{\alpha }{r}+\sum_{i=1}^{n}\frac{\beta _{i}}{%
2x_{i}^{2}}\qquad .  \label{V_CR}
\end{equation}

We shall call the system governed by the Hamiltonian given above the
Coulomb-Rosochatius\cite{Ro77} system.The motion is confined to the region
bounded by the singularity of $H$\ at hyper-planes $x_{i}=0,\ i=1,2,\ldots
,n $\ ,\ i.e. hyper-octant ( one of the $2^{n}$\ regions of Euclidean space $%
E^{n}$). Without loss of generality this can be taken to be the non-negative
hyper-octant ( i.e. where $x_{i}\geq 0,\ i=1,2,\ldots ,n$\ ).

The main interest of the Coulomb-Rosochatius system consists in its maximal
superintegrability, recently proved for $n=3$\ in the classical case\cite%
{VE08} (see also \cite{Ro09}). The system had been proved long ago \cite%
{Ma67},\cite{Ev90} to be quasi-maximally superintegrable for $n\leq 3$, i.e.
to admit $2n-2$\ operators quadratic in the momenta. For $n=3$, the fifth
integral of motion quartic in the momenta was explicitly worked out in Refs%
\cite{VE08}-\cite{Ro09}, thus proving the maximal superintegrability of the
classical Coulomb-Rosochatius system. Finally, the maximal
superintegrability of the classical system (\ref{V_CR}) in $n$-dimensional
spherical, hyperbolic and Euclidean spaces was proved in Ref.\cite{BH09}:
here again, one of the $2n-1$\ functionally independent integrals of motion
turns out to be quartic in the momenta, while the remaining ones are
quadratic.

The main purpose of the present work is the complete algebraic solution of
bound states and scattering states of the quantum-mechanical
Coulomb-Rosochatius system in $n$ dimensions and the derivation of the $2n-1$
integrals of motion expected for maximal superintegrability.

Before describing the method of solution of the quantum Coulomb-Rosochatius
system in $n$ dimensions , a few definitions are in order: a Lie group $G$
is an invariance group for a quantum-mechanical system with Hamiltonian $H$
if the latter can be related to a suitable function of a Casimir operator, $%
C $, of the group, $G$ 
\begin{equation}
H=f\left( C\right) \;.  \label{inv_group}
\end{equation}
An algebraic derivation of the energy spectrum is possible also when 
\begin{equation}
H=f\left( C\right) |_{\mathcal{H}}.  \label{pot_group}
\end{equation}
where $\mathcal{H}$ is a subspace of the carrier space. In this case the
group $G$\ describes the same energy states of a family of Hamiltonians $H$\
with different potential strength. This is why $G$\ is called potential
group \cite{AGI83}. Such an approach was proposed by Ghirardi \cite{Gh72},
who\ worked it out in detail for the Scarf potential\ \cite{Sc58}. It is
similar to the approach of Olshanetsky and Perelomov \cite{OP77, OP83},
where quantum integrable systems are related to the radial part of the
Laplace operator on homogeneous spaces,i.e$.$\ to the radial part of a
second-order Casimir operator of a Lie group. Moreover, it has been shown in
Ref. \cite{Ke98} that the $S$ matrix \ can be associated with an
intertwining operator, $A$, between two Weyl-equivalent representations, $%
U^{\chi }$ and $U^{\widetilde{\chi }}$, of $G$, \textit{i.e.} two
representations with the same Casimir eigenvalues.

By definition, $A$ satisfies the following equations 
\begin{equation}
AU^{\chi }\left( g\right) =U^{\widetilde{\chi }}\left( g\right)
A\;,\;\forall \;g\;\epsilon \;G  \label{intertw_1}
\end{equation}
and 
\begin{equation}
AdU^{\chi }\left( b\right) =dU^{\widetilde{\chi }}\left( b\right) \;,\;\
\forall \;b\;\epsilon \;\mathfrak{g}  \label{intertw_2}
\end{equation}
where $dU^{\chi }$ and $dU^{\widetilde{\chi }}$ are the corresponding
representations of the algebra, $\mathfrak{g}$, of $G$. Eqs. (\ref{intertw_1}%
-\ref{intertw_2}) have high restrictive power, determining the intertwining
operator up to a constant. The $S$ matrix coincides with the intertwining
operator $A$%
\begin{equation}
S=A  \label{S_A_1}
\end{equation}
if eq. (\ref{inv_group}) holds, and with the reduction of $A$ to a proper
Hilbert subspace $\mathcal{H}$%
\begin{equation}
S=A|_{\mathcal{H}}  \label{S_A_2}
\end{equation}
if eq. (\ref{pot_group}) holds.

The plan of the paper is as follows: Section 2 will describe in full detail
the potential group for the Coulomb-Rosochatius system, with a potential
specialized to 
\begin{equation}
V(r)=-\frac{\gamma }{r}+\sum_{i=1}^{n}\frac{\kappa _{i}\left( \kappa
_{i}+1\right) }{x_{i}^{2}},  \label{non_centr_V2}
\end{equation}
with Subsection 2.1 dedicated to the derivation of bound states and
Subsection 2.2 to scattering states. Finally, Section 3 will be devoted to
conclusions and perspectives.

\section{Potential group for the Coulomb-Rosochatius system}

It is well known that the group $SO\left( N+1\right) $ ( $SO\left(
N,1\right) $ ) has a class of unitary irreducible representations (UIR's)
characterized by a number $j,\ j=0,1,2,\ldots $\ ($j=-\frac{N-1}{2}+i\rho ,\
\rho >0$) in which the basis vectors are completely labelled by $N-1$\
numbers only. This representation can be realized in the Hilbert space
spanned by negative-energy (positive-energy) states corresponding to a fixed
eigenvalue of the Coulomb Hamiltonian in $N$ dimensions. According to this,
one introduces the angular momentum operators $L_{ij}$ and the $N$%
-dimensional Runge-Lenz vector $A_{i}$ 
\begin{equation}
L_{ij}=\zeta _{i}\pi _{j}-\zeta _{j}\pi _{i}\;,\;\left( i,j=1,...,N\right)
\end{equation}
and 
\begin{eqnarray}
A_{i} &=&-\frac{1}{2}\sum_{j=1}^{N}\left( L_{ij}\pi _{j}+\pi
_{j}L_{ij}\right) +\frac{\gamma \zeta _{i}}{\sqrt{\zeta ^{2}}}\;  \nonumber
\\
&=&\left( \zeta \cdot \pi \right) \pi _{i}-\zeta _{i}\pi ^{2}-i\frac{N-1}{2}%
\pi _{i}+\frac{\gamma \zeta _{i}}{\sqrt{\zeta ^{2}}},\;(i=1,...,N)
\end{eqnarray}
respectively, where $\zeta =\left( \zeta _{1},\zeta _{2},\ldots ,\zeta
_{N}\right) $, $\ \ \pi =\left( \ \pi _{1},\pi _{2},\ldots ,\pi _{N}\right)
,\ \pi _{j}=-i\frac{\partial }{\partial \zeta _{j}}$, , $\zeta
^{2}=\sum_{i=1}^{N}\zeta _{i}^{2}$, $\pi ^{2}=\sum_{i=1}^{N}\pi _{i}^{2}$.
These operators satisfy the following commutation relations 
\[
\left[ L_{ij},L_{kl}\right] ={\normalsize i}\left( {\normalsize \delta }%
_{ik}L_{jl}+{\normalsize \delta }_{jl}L_{ik}-{\normalsize \delta }%
_{il}L_{jk}-{\normalsize \delta }_{jk}L_{il}\right) 
\]
\begin{equation}
\left[ L_{ij},A_{k}\right] ={\normalsize i}\left( {\normalsize \delta }%
_{ik}A_{j}-{\normalsize \delta }_{jk}A_{i}\right)
\end{equation}
\[
\left[ A_{i},A_{j}\right] =-{\normalsize i}2hL_{ij}\;, 
\]
where $h$\ is the Coulomb Hamiltonian in $N$ dimensions 
\begin{equation}
h=\frac{1}{2}\pi ^{2}-\frac{\gamma }{\sqrt{\zeta ^{2}}}.  \label{C_H}
\end{equation}

Since $\left[ L_{ij},h\right] =\left[ A_{i},h\right] =0$\ , we may restrict
the above algebra to a subspace where $h$ has a definite eigenvalue $%
\varepsilon $\ and define a new set of operators as follows 
\begin{eqnarray}
M_{i,0} &=&-M_{0,i}=\left| 2\varepsilon \right| ^{-\frac{1}{2}%
}A_{i}\;,\;(i=1,...,N)  \nonumber \\
M_{ij} &=&L_{ij}\;,\;\left( i,j=1,...,N\right) .  \label{Rep}
\end{eqnarray}
Then, as a result, we obtain the Lie algebra of $SO\left( N+1\right) $ (when 
$\varepsilon $\ is negative) and the Lie algebra of $SO\left( N,1\right) $
(when $\varepsilon $\ is positive) 
\begin{equation}
\left[ M_{\mu \nu },M_{\sigma \lambda }\right] ={\normalsize i}\left( g_{\mu
\sigma }M_{\nu \lambda }+g_{\nu \lambda }M_{\mu \sigma }-g_{\mu \lambda
}M_{\nu \sigma }-g_{\nu \sigma }M_{\mu \lambda }\right) \;,  \label{Comm_rel}
\end{equation}
where $\mu ,\nu =0,1,2,...,N$\ and 
\begin{eqnarray}
g_{\mu \nu } &=&\left( +,+,\ldots ,+,+\right) \quad \textstyle{for \ \ }%
SO\left( N+1\right) \\
g_{\mu \nu } &=&\left( -,+,\ldots ,+,+\right) \quad \textstyle{for \ \ }%
SO\left( N,1\right)  \nonumber
\end{eqnarray}
The generators $M_{\mu \nu }$ act in the eigenspace $\mathcal{H}$ of $h$\
equipped with the scalar product 
\begin{equation}
\left( \phi _{1},\phi _{2}\right) =\int\limits_{R^{N}}\phi _{1}^{\ast
}\left( \zeta \right) \phi _{2}\left( \zeta \right) d\zeta ,\quad \zeta \in
R^{N}
\end{equation}
where $d\zeta =d\zeta _{1}d\zeta _{2}\cdots d\zeta _{N}$. A detailed
discussion of the $SO\left( N+1\right) $ ($SO\left( N,1\right) $)
representation \ generated by operators (\ref{Rep}) is given in Refs.\cite%
{Al58}-\cite{RS79}.

At this stage we note that, in \ general, one can define the generators of $%
SO\left( N+1\right) $ ( $SO\left( N,1\right) $ ) (let us call them $\tilde{M}%
_{\mu \nu }$), as follows 
\begin{eqnarray*}
\tilde{M}_{i,0} &=&-\tilde{M}_{0,i}=\left| 2\varepsilon \right| ^{-\frac{1}{2%
}}\lambda ^{1/2}\left( \zeta \right) \circ A_{i}\circ \lambda ^{-1/2}\left(
\zeta \right) \\
\tilde{M}_{ij} &=&\lambda ^{1/2}\left( \zeta \right) \circ L_{ij}\circ
\lambda ^{-1/2}\left( \zeta \right)
\end{eqnarray*}
where $\lambda $ is some non-negative function of $\ \zeta $. Now the
generators $\tilde{M}_{\mu \nu }$ act in the eigenspace $\tilde{\mathcal{H}}$%
\ of $\tilde{h}=\lambda ^{1/2}\left( \zeta \right) \circ h\circ \lambda
^{-1/2}\left( \zeta \right) $\ equipped with the scalar product 
\begin{equation}
\left( \tilde{\phi}_{1},\tilde{\phi}_{2}\right) =\int\limits_{R^{N}}\tilde{%
\phi}_{1}^{\ast }\left( \zeta \right) \tilde{\phi}_{2}\left( \zeta \right)
d\mu \left( \zeta \right) ,\quad \zeta \in R^{N}
\end{equation}
where $d\mu \left( \zeta \right) =\lambda ^{-1}\left( \zeta \right) d\zeta $
is a quasi-invariant measure on \ $R^{N}$. The representations acting in $%
\mathcal{H}$ and $\tilde{\mathcal{H}}$ are, of course, unitarily equivalent.
The unitary mapping $W$ which realizes the equivalence is given by 
\begin{equation}
W:\textstyle{\ \ \ \ }\phi \rightarrow \tilde{\phi}=\lambda ^{1/2}\left(
\zeta \right) \phi  \label{Map}
\end{equation}

Although these representations are equivalent from the mathematical
view-point, they may be related to different physical problems. We shall
prove that the bound states and scattering states of the quantum system (\ref%
{non_centr_V2}) are related to the representation of $SO(3n+1)$ and $%
SO(3n,1) $ , respectively, acting in the Hilbert space with scalar product 
\begin{equation}
\left( \tilde{\phi}_{1},\tilde{\phi}_{2}\right) =\int\limits_{R^{3n}}\tilde{%
\phi}_{1}^{\ast }\left( \zeta \right) \tilde{\phi}_{2}\left( \zeta \right)
d\mu \left( \zeta \right) ,  \label{scalar}
\end{equation}%
where $d\mu \left( \zeta \right) =\prod_{i=1}^{n}\left( \zeta
_{3i-2}^{2}+\zeta _{3i-1}^{2}+\zeta _{3i}^{2}\right) ^{-1}d\zeta $.\ To this
end, we choose $N=3n$ and $\lambda \left( \zeta \right)
=\prod_{i=1}^{n}\left( \zeta _{3i-2}^{2}+\zeta _{3i-1}^{2}+\zeta
_{3i}^{2}\right) $. Then we introduce the second-order Casimir operator 
\begin{equation}
\tilde{C}=\frac{1}{2}\sum\limits_{\mu ,\nu =0}^{3n}\tilde{M}^{\mu \nu }%
\tilde{M}_{\mu \nu }  \label{Cas_op}
\end{equation}%
and consider the following operator depending on $\tilde{C}$%
\begin{eqnarray*}
&&\frac{\gamma ^{2}}{\left[ \tilde{C}+\left( \frac{3n-1}{2}\right) ^{2}%
\right] } \\
&=&\frac{\partial ^{2}}{\partial \zeta _{1}^{2}}+\ldots +\frac{\partial ^{2}%
}{\partial \zeta _{3n}^{2}}-\frac{2}{\zeta _{1}^{2}+\zeta _{2}^{2}+\zeta
_{3}^{2}}\left( \zeta _{1}\frac{\partial }{\partial \zeta _{1}}+\zeta _{2}%
\frac{\partial }{\partial \zeta _{2}}+\zeta _{3}\frac{\partial }{\partial
\zeta _{3}}\right) \\
&&-\cdots -\frac{2}{\zeta _{3n-2}^{2}+\zeta _{3n-1}^{2}+\zeta _{3n}^{2}}%
\left( \zeta _{3n-2}\frac{\partial }{\partial \zeta _{3n-2}}+\zeta _{3n-1}%
\frac{\partial }{\partial \zeta _{3n-1}}+\zeta _{3n}\frac{\partial }{%
\partial \zeta _{3n}}\right) +\frac{2\gamma }{\sqrt{\zeta ^{2}}}
\end{eqnarray*}%
Decomposing $R^{3n}$\ into the direct sum of three-dimensional subspaces and
introducing spherical coordinates in these subspaces 
\begin{equation}
\zeta =\left( x_{1}u_{1},x_{2}u_{2},\ldots ,x_{n}u_{n}\right)
\label{zeta_vec}
\end{equation}%
where $u_{i}=\left( \sin \alpha _{i}\sin \beta _{i},\sin \alpha _{i}\cos
\beta _{i},\cos \alpha _{i}\right) ,\ x_{i}\geq 0,\ 0\leq \alpha _{i}\leq
\pi ,\ 0\leq \beta _{i}\leq 2\pi ,\ (i=1,2,\ldots ,n)$ , we have 
\begin{eqnarray}
&&\frac{\gamma ^{2}}{\left[ \tilde{C}+\left( \frac{3n-1}{2}\right) ^{2}%
\right] }  \label{C_dep_op} \\
&=&\sum_{i=1}^{n}\left[ \frac{\partial ^{2}}{\partial x_{i}^{2}}+\frac{1}{%
x_{i}^{2}}\left( \frac{1}{\sin \alpha _{i}}\frac{\partial }{\partial \alpha
_{i}}\sin \alpha _{i}\frac{\partial }{\partial \alpha _{i}}+\frac{1}{\sin
^{2}\alpha _{i}}\frac{\partial ^{2}}{\partial \beta _{i}^{2}}\right) \right]
+\frac{2\gamma }{\sqrt{x^{2}}}  \nonumber
\end{eqnarray}%
with $x=\left( x_{1},x_{2},\ldots ,x_{n}\right) .$ With this coordinate
system the measure $d\mu \left( \zeta \right) $ in formula (\ref{scalar})
becomes $d\mu \left( \zeta \right) =dx\prod_{i=1}^{n}\sin \alpha _{i}d\alpha
_{i}d\beta _{i}$, where $dx=dx_{1}dx_{2}\cdots dx_{n}.$

Let $\mathcal{H}_{K}$, $K=\left( \kappa _{1},\kappa _{2},\ldots ,\kappa
_{n};\sigma _{1},\sigma _{2},\ldots ,\sigma _{n}\right) $, be a subspace of
functions $\hat{\phi}\left( \zeta \right) =\Psi \left( x\right)
\prod_{i=1}^{n}Y_{\kappa _{i}}^{\sigma _{i}}\left( \alpha _{i},\beta
_{i}\right) ,$ with fixed $K$, where $Y_{\kappa _{i}}^{\sigma _{i}}\left(
\alpha _{i},\beta _{i}\right) $ are spherical harmonics of degree $\kappa
_{i}$. Thus, the operator (\ref{C_dep_op}) restricted to this subspace
becomes 
\begin{equation}
\left. \frac{\gamma ^{2}}{\left[ \tilde{C}+\left( \frac{3n-1}{2}\right) ^{2}%
\right] }\right| _{\mathcal{H}_{K}}=\sum_{i=1}^{n}\left[ \frac{\partial ^{2}%
}{\partial x_{i}^{2}}-\frac{\kappa _{i}\left( \kappa _{i}+1\right) }{%
x_{i}^{2}}\right] +\frac{2\gamma }{\sqrt{x^{2}}}\;,
\end{equation}
where we have used the fact that 
\[
\left( \frac{1}{\sin \alpha _{i}}\frac{\partial }{\partial \alpha _{i}}\sin
\alpha _{i}\frac{\partial }{\partial \alpha _{i}}+\frac{1}{\sin ^{2}\alpha
_{i}}\frac{\partial ^{2}}{\partial \beta _{i}^{2}}\right) Y_{\kappa
_{i}}^{\sigma _{i}}\left( \alpha _{i},\beta _{i}\right) =-\kappa _{i}\left(
\kappa _{i}+1\right) Y_{\kappa _{i}}^{\sigma _{i}}\left( \alpha _{i},\beta
_{i}\right) 
\]
Hence, the Hamiltonian 
\begin{equation}
H=-\frac{1}{2}\nabla ^{2}-\frac{\gamma }{\sqrt{x^{2}}}+\sum_{i=1}^{n}\frac{%
\kappa _{i}\left( \kappa _{i}+1\right) }{2x_{i}^{2}}  \nonumber
\end{equation}
can be described in terms of the potential groups $SO\left( 3n+1\right) $
and $SO\left( 3n,1\right) $ since 
\[
H=-\left. \frac{\gamma ^{2}}{2\left[ \tilde{C}+\left( \frac{3n-1}{2}\right)
^{2}\right] }\right| _{\mathcal{H}_{K}}\;, 
\]
Moreover, since the Casimir operators of the groups in the chain 
\begin{eqnarray}
G &\supset &SO\left( 3n\right) \supset SO\left( 3n-3\right) \times SO\left(
3\right) \supset SO\left( 3n-6\right) \times SO\left( 3\right) \times
SO\left( 3\right) \supset \ldots  \nonumber \\
&\supset &SO\left( 3\right) \times SO\left( 3\right) \times \ldots \times
SO\left( 3\right) \supset SO\left( 2\right) \times SO\left( 2\right) \times
\ldots \times SO\left( 2\right)  \label{chain}
\end{eqnarray}
where $G$ is either $SO\left( 3n+1\right) $ or $SO\left( 3n,1\right) $, form
a complete set of $3n$ commuting operators ( including the Casimir operator, 
$\widetilde{C}$, of $G$\ ), i.e. the operators $\widetilde{C},\ \tilde{C}%
^{SO(3n-3p)},\ \tilde{C}^{SO(3)_{k}}$\ and $\tilde{C}^{SO(2)_{k}},$\ where $%
\ p=0,1,2,\ldots ,n-2,\ k=1,2,\ldots ,n$ and\ 
\begin{eqnarray}
\tilde{C}^{SO(3n-3p)} &=&\frac{1}{2}\sum_{i,j=1}^{3n-3p}\tilde{M}_{ij}^{2}=-%
\frac{1}{2}\sum_{i,j=1}^{n-p}\left( x_{i}\frac{\partial }{\partial x_{j}}%
-x_{j}\frac{\partial }{\partial x_{i}}\right) ^{2}-\left(
\sum_{i=1}^{n-p}x_{i}^{2}\right)  \label{Cas_1} \\
&&\times \sum_{j=1}^{n-p}\frac{1}{x_{j}^{2}}\left[ \frac{1}{\sin \alpha _{j}}%
\frac{\partial }{\partial \alpha _{j}}\sin \alpha _{j}\frac{\partial }{%
\partial \alpha _{j}}+\frac{1}{\sin ^{2}\alpha _{j}}\frac{\partial ^{2}}{%
\partial \beta _{j}^{2}}\right] -2\left( n-p\right) \left( n-p-1\right) , 
\nonumber
\end{eqnarray}

\begin{equation}
\tilde{C}^{SO(3)_{k}}=\frac{1}{2}\sum_{i,j=3\left( n-k\right) +1}^{3\left(
n-k\right) +3}\tilde{M}_{ij}^{2}=-\left( \frac{1}{\sin \alpha _{k}}\frac{%
\partial }{\partial \alpha _{k}}\sin \alpha _{k}\frac{\partial }{\partial
\alpha _{k}}+\frac{1}{\sin ^{2}\alpha _{k}}\frac{\partial ^{2}}{\partial
\beta _{k}^{2}}\right) ,  \label{Cas_SO3}
\end{equation}
and 
\begin{equation}
\tilde{C}^{SO(2)_{k}}=\frac{1}{2}\sum_{i,j=3\left( n-k\right) +2}^{3\left(
n-k\right) +3}\tilde{M}_{ij}^{2}=-\frac{\partial ^{2}}{\partial \beta
_{k}^{2}}  \label{Cas_SO2}
\end{equation}
are mutually commuting operators, it follows from formula (\ref{Cas_1}) that
the $n-1$ operators 
\begin{eqnarray}
I_{p} &=&\left. \left[ \tilde{C}^{SO(3n-3p)}+2\left( n-p\right) \left(
n-p-1\right) \right] \right| _{\mathcal{H}_{K}}  \label{I_p} \\
&=&-\frac{1}{2}\sum_{i,j=1}^{n-p}\left( x_{i}\frac{\partial }{\partial x_{j}}%
-x_{j}\frac{\partial }{\partial x_{i}}\right)
^{2}-\sum_{i=1}^{n-p}x_{i}^{2}\sum_{j=1}^{n-p}\frac{\kappa _{j}\left( \kappa
_{j}+1\right) }{x_{j}^{2}},\ \;p=0,1,2,\ldots ,n-2  \nonumber
\end{eqnarray}
are integrals of motion. These integrals of motion are responsible for the
separability of $H$ in spherical coordinates.

The remaining $n-1$ integrals of motion can be related to Casimir operators 
\begin{eqnarray*}
\tilde{C}^{SO(3n-3q)} &=&\frac{1}{2}\sum_{i,j=3q+1}^{3n}\tilde{M}_{ij}^{2}=-%
\frac{1}{2}\sum_{i,j=q+1}^{n}\left( x_{i}\frac{\partial }{\partial x_{j}}%
-x_{j}\frac{\partial }{\partial x_{i}}\right) ^{2}-\left(
\sum_{i=q+1}^{n}x_{i}^{2}\right) \\
&&\times \sum_{j=q+1}^{n}\frac{1}{x_{j}^{2}}\left[ \frac{1}{\sin \alpha _{j}}%
\frac{\partial }{\partial \alpha _{j}}\sin \alpha _{j}\frac{\partial }{%
\partial \alpha _{j}}+\frac{1}{\sin ^{2}\alpha _{j}}\frac{\partial ^{2}}{%
\partial \beta _{j}^{2}}\right] -2\left( n-q\right) \left( n-q-1\right) ,
\end{eqnarray*}%
\begin{eqnarray*}
\tilde{C}^{G^{\prime }} &=&\frac{1}{2}\sum\limits_{\mu ,\nu =0}^{3}\tilde{M}%
^{\mu \nu }\tilde{M}_{\mu \nu }=-\frac{1}{2\varepsilon }\left\{ \left[ 
\mathcal{A}_{1}+x_{1}\sum_{i=1}^{n}\frac{1}{x_{i}^{2}}\left( \frac{1}{\sin
\alpha _{i}}\frac{\partial }{\partial \alpha _{i}}\sin \alpha _{i}\frac{%
\partial }{\partial \alpha _{i}}+\frac{1}{\sin ^{2}\alpha _{i}}\frac{%
\partial ^{2}}{\partial \beta _{i}^{2}}\right) \right] ^{2}\right. \\
&&\left. -\left( p\cdot x\frac{1}{x_{1}^{2}}x\cdot p-2\varepsilon \right)
\left( \frac{1}{\sin \alpha _{1}}\frac{\partial }{\partial \alpha _{1}}\sin
\alpha _{1}\frac{\partial }{\partial \alpha _{1}}+\frac{1}{\sin ^{2}\alpha
_{1}}\frac{\partial ^{2}}{\partial \beta _{1}^{2}}\right) -\frac{\left(
n-1\right) \left( n-3\right) }{4x_{1}^{2}}+2\varepsilon \right\}
\end{eqnarray*}%
with $\mathcal{A}_{1}=\left( x\cdot p\right) p_{1}-x_{1}p^{2}-i\frac{n-1}{2}%
p_{1}+\frac{\gamma x_{1}}{\sqrt{x^{2}}},$ $x=\left( x_{1},x_{2},\ldots
,x_{n}\right) $$\ \ p=\left( \ p_{1},p_{2},\ldots ,p_{n}\right) $ and $\
p_{j}=-i\frac{\partial }{\partial x_{j}}$, of the groups in the chain 
\begin{eqnarray*}
G &\supset &SO\left( 3n-3\right) \times G^{\prime }\supset SO\left(
3n-6\right) \times SO\left( 3\right) \times SO\left( 3\right) \supset \ldots
\\
&\supset &SO\left( 3\right) \times SO\left( 3\right) \times \ldots \times
SO\left( 3\right) \supset SO\left( 2\right) \times SO\left( 2\right) \times
\ldots \times SO\left( 2\right)
\end{eqnarray*}

where $G^{\prime }$ is either $SO\left( 4\right) $ or $SO\left( 3,1\right) $%
. Namely, 
\begin{eqnarray}
J_{q} &=&\left. \left[ \tilde{C}^{SO(3n-3q)}+2\left( n-q\right) \left(
n-q-1\right) \right] \right\vert _{\mathcal{H}_{K}}  \label{J_q} \\
&=&-\frac{1}{2}\sum_{i,j=q+1}^{n}\left( x_{i}\frac{\partial }{\partial x_{j}}%
-x_{j}\frac{\partial }{\partial x_{i}}\right)
^{2}-\sum_{i=q+1}^{n}x_{i}^{2}\sum_{j=q+1}^{n}\frac{\kappa _{j}\left( \kappa
_{j}+1\right) }{x_{j}^{2}},\ q=1,2,\ldots ,n-2  \nonumber
\end{eqnarray}%
and

\begin{eqnarray}
J_{n-1} &=&-2\varepsilon \left. \left[ \tilde{C}^{G^{\prime }}-\tilde{C}%
^{SO(3)_{1}}-1\right] \right| _{\mathcal{H}_{K}}  \label{Jl} \\
&=&\left[ \mathcal{A}_{1}-x_{1}\sum_{i=1}^{n}\frac{\kappa _{i}\left( \kappa
_{i}+1\right) }{x_{i}^{2}}\right] ^{2}+\kappa _{1}\left( \kappa
_{1}+1\right) \left( p\cdot x\right) \frac{1}{x_{1}^{2}}\left( x\cdot
p\right) -\frac{\left( n-1\right) \left( n-3\right) }{4x_{1}^{2}}  \nonumber
\end{eqnarray}
are constants of motion, too.

It is worth noting that, since $J_{n-1}$ is quartic in the momenta, the
complete set of mutually commuting operators $\left\{ H,J_{q},\ q=1,2,\ldots
,n-1\right\} $ does not specify a separable coordinate system.

\subsection{Bound states}

The bound-state spectrum is immediately obtained from the eigenvalue of the
Casimir operator $\tilde{C}$ of the potential group $SO\left( 3n+1\right) $, 
\textit{i.e.} $j\left( j+3n-1\right) $, in the form 
\begin{equation}
E=-\frac{\gamma ^{2}}{2\left( j+\frac{3n-1}{2}\right) ^{2}}\;,  \label{E_b}
\end{equation}
where $j$\ takes on integer values from $\kappa _{1}+\kappa _{2}+\ldots
+\kappa _{n}$ upwards.

The basis functions $\left| j;lMK\right\rangle $ for Hilbert space $\tilde{%
\mathcal{H}}$ can be defined as the common set of eigenfunctions of  the
Casimir operators of \ the groups forming the chain (\ref{chain}) 
\begin{eqnarray*}
\tilde{C}\left| j;lMK\right\rangle &=&j\left( j+3n-1\right) \left|
j;lMK\right\rangle \\
\tilde{C}^{SO(3n-3p)}\left| j;lMK\right\rangle &=&m_{p}\left(
m_{p}+3n-3p-2\right) \left| j;lMK\right\rangle ,\quad p=0,1,\ldots ,n-2 \\
\tilde{C}^{SO(3)_{i}}\left| j;lMK\right\rangle &=&\kappa _{i}\left( \kappa
_{i}+1\right) \left| j;lMK\right\rangle ,\ \ i=1,2,\ldots ,n \\
\tilde{C}^{SO(2)_{i}}\left| j;lMK\right\rangle &=&\sigma _{i}^{2}\left|
j;lMK\right\rangle ,\ \ i=1,2,\ldots ,n
\end{eqnarray*}
where $m_{0}\equiv l$ and $M$\ and $K$\ are the collective indexes$\ \left(
m_{1},m_{2},\ldots ,m_{n-2}\right) $ and \ $K=\left( \kappa _{1},\kappa
_{2},\ldots ,\kappa _{n};\sigma _{1},\sigma _{2},\ldots ,\sigma _{n}\right) $%
, respectively. It is also known that in correspondence with every chain of
subgroups in $SO(N)$ it is possible to define a polyspherical system on $%
R^{N}$ (see Ch. 10 of Ref.\cite{VK93}). According to this, we introduce
polyspherical coordinates on $R^{3n}$\ by choosing $x_{i}$\ in (\ref%
{zeta_vec}) as

\begin{equation}
\begin{array}[t]{l}
x_{1}=r\sin \theta _{n-1}\sin \theta _{n-2}\ldots \sin \theta _{2}\sin
\theta _{1} \\ 
x_{2}=r\sin \theta _{n-1}\sin \theta _{n-2}\ldots \sin \theta _{2}\cos
\theta _{1} \\ 
\cdots \\ 
x_{n-1}=r\sin \theta _{n-1}\cos \theta _{n-2} \\ 
x_{n}=r\cos \theta _{n-1}%
\end{array}
\label{x_vec}
\end{equation}
where $r\geq 0,\ 0\leq \theta _{i}\leq \pi /2,\ $ for $i=1,2,\ldots ,n-1$.

By construction 
\[
\ \left\langle \zeta \right. \left\vert j;lMK\right\rangle =\psi \left(
x\right) \prod_{i=1}^{n}Y_{\kappa _{i}}^{\sigma _{i}}\left( \alpha
_{i},\beta _{i}\right) 
\]
where $\psi \left( x\right) $\ is the bound-state wave function 
\begin{equation}
\psi \left( x\right) =\mathcal{R}_{jl}\left( r\right) \mathcal{Y}_{lM}\left( 
\hat{x}\right) ,\ \hat{x}=x/r  \label{psi_b}
\end{equation}
Here $\mathcal{R}_{jl}\left( r\right) $ is the radial part of the wave
function, while $\mathcal{Y}_{lM}\left( \hat{x}\right) $ is the angular part
of it. According to (\ref{Map}) $\mathcal{R}_{jl}\left( r\right) $\ is
related to the radial part $\mathcal{R}_{jl}^{Coul}\left( r\right) $\ of the 
$3n$-dimensional Coulomb wave function \cite{Ni79} as 
\begin{eqnarray*}
\mathcal{R}_{jl}\left( r\right) &=&r^{n}\mathcal{R}_{jl}^{Coul}\left(
r\right) \\
&=&cu^{l+n}e^{-\frac{u}{2}}L_{j-l}^{2l+3n-2}\left( u\right) \;,\;u=4\gamma
r/\left( 2j+3n-1\right) \;,
\end{eqnarray*}
\ where \ $L_{n}^{\alpha }$ are Laguerre polynomials and 
\begin{equation}
c=\left( 2\gamma \right) ^{-n/2}\left[ j+\frac{1}{2}\left( 3n-1\right) %
\right] ^{-\frac{1}{2}\left( n+1\right) }\left[ \frac{\Gamma \left(
j-l+1\right) }{2\Gamma \left( j+l+3n-1\right) }\right] ^{\frac{1}{2}}\;,
\label{c}
\end{equation}
while $\mathcal{Y}_{lM}\left( \hat{x}\right) $ are related to the ($3n-1)$%
-dimensional spherical harmonics $Y_{lMK}\left( \hat{\zeta}\right) $\ in the
above polyspherical coordinates as 
\begin{equation}
Y_{lMK}\left( \hat{\zeta}\right) =\mathcal{Y}_{lM}\left( \hat{x}\right)
\prod_{k=1}^{n-1}\left( \sin ^{n-k}\theta _{n-k}\cos \theta _{n-k}\right)
^{-1}\prod_{i=1}^{n}Y_{\kappa _{i}}^{\sigma _{i}}\left( \alpha _{i},\beta
_{i}\right) ,\ \hat{\zeta}=\zeta /r  \label{Harm}
\end{equation}
\qquad 
\begin{figure}[tbp]
\begin{center}
\includegraphics[width=12cm,angle=0,clip]{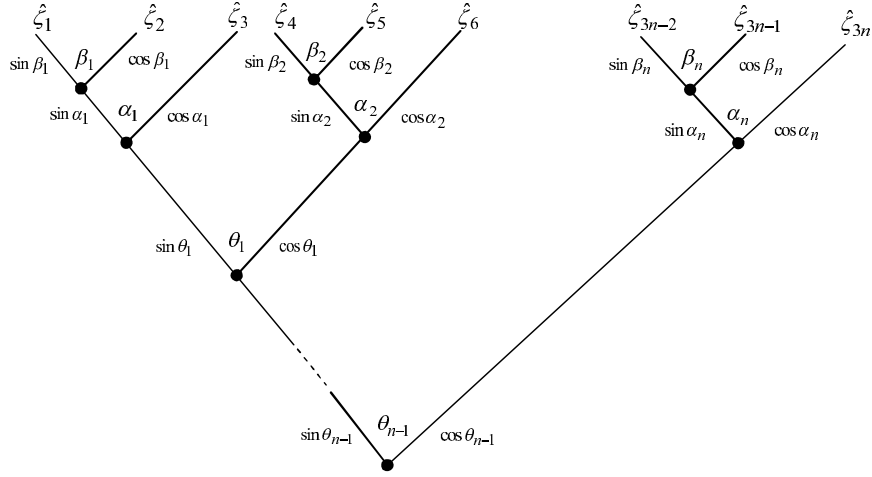}
\end{center}
\caption{A tree of polyspherical coordinates for $S^{3n-1}$.}
\label{Fig1}
\end{figure}
In order to write $Y_{lMK}\left( \hat{\zeta}\right) $\ we make use of a
graphical method called the tree method (see Section 10.5 of Ref.\cite{VK93}%
). According to this method polyspherical coordinate systems on the sphere
of unit radius can be described by graphs called trees. The trees contain
nodes and each node has two edges, which are distinguished as left or right
edge. An angle is associated with each node and the sine (cosine) of this
angle is associated with the corresponding left (right) edge. The free
(upper) ends of tree are labelled left to right with Cartesian coordinates.
Then the Cartesian coordinate $\hat{\zeta}_{i}$\ is equal to the product of
trigonometrical functions on the edges along the unique path connecting the
lowest node with \ $\hat{\zeta}_{i}$. In Fig.1 we associate a tree with a
polyspherical coordinate system on $S^{3n-1}$\ given by formulae (\ref%
{zeta_vec}) and (\ref{x_vec}). Hence, the ($3n-1)$-dimensional spherical
harmonics $Y_{lMK}\left( \hat{\zeta}\right) $ corresponding to this tree can
be obtained by the rules described in Section 10.5.3 of Ref.\cite{VK93}. As
a result we have 
\begin{eqnarray}
\mathcal{Y}_{lM}\left( \hat{x}\right) &=&\chi \prod_{i=1}^{n-2}\sin
^{m_{i}+n-i}\theta _{n-i}\cos ^{\kappa _{i}+1}\theta _{n-i}P_{\left(
m_{i-1}-m_{i}-\kappa _{i}\right) /2}^{\left( m_{i}+\frac{3n-3i}{2}-1,\quad
\kappa _{i}+\frac{1}{2}\right) }\left( \cos 2\theta _{n-i}\right)  \nonumber
\\
&&\times \sin ^{\kappa _{n}+1}\theta _{1}\cos ^{\kappa _{n-1}+1}\theta
_{1}P_{\left( m_{n-2}-\kappa _{n}-\kappa _{n-1}\right) /2}^{\left( \kappa
_{n}+\frac{1}{2},\quad \kappa _{n-1}+\frac{1}{2}\right) }\left( \cos 2\theta
_{1}\right) \;,
\end{eqnarray}
where $P_{n}^{\left( \alpha ,\beta \right) }$ are Jacobi polynomials and $%
\chi $ is a normalization constant 
\begin{eqnarray}
\chi &=&\prod_{i=1}^{n-2}\left[ \frac{\Gamma \left( \frac{1}{2}\left(
m_{i-1}+m_{i}+\kappa _{i}+3n-3i+1\right) \right) \Gamma \left( \frac{1}{2}%
\left( m_{i-1}-m_{i}-\kappa _{i}+2\right) \right) \left(
2m_{i}+3n-3i+1\right) }{\Gamma \left( \frac{1}{2}\left( m_{i-1}+m_{i}-\kappa
_{i}\right) \right) \Gamma \left( \frac{1}{2}\left( m_{i-1}-m_{i}+\kappa
_{i}+3\right) \right) }\right] ^{\frac{1}{2}}  \nonumber \\
&&\times \left[ \frac{\Gamma \left( \frac{1}{2}\left( m_{n-2}+\kappa
_{n}+\kappa _{n-1}+4\right) \right) \Gamma \left( \frac{1}{2}\left(
m_{n-2}-\kappa _{n}-\kappa _{n-1}+2\right) \right) \left( 2m_{n-2}+4\right) 
}{\Gamma \left( \frac{1}{2}\left( m_{n-2}+\kappa _{n}-\kappa _{n-1}+3\right)
\right) \Gamma \left( \frac{1}{2}\left( m_{n-2}-\kappa _{n}+\kappa
_{n-1}+3\right) \right) }\right] ^{\frac{1}{2}}  \label{chi}
\end{eqnarray}

\subsection{Scattering states}

Once the group structure of the problem has been recognized, the associated $%
S$ matrix can be computed by using matrices \ that intertwine
Weyl-equivalent representations of $SO\left( 3n,1\right) $ in the bases
corresponding to the reduction (\ref{chain}). We find it expedient to use,
for this purpose, equation (\ref{intertw_1}). By realizing the principal
series of $SO\left( 3n,1\right) $ on suitable Hilbert spaces of appropriate
functions, one can derive from (\ref{intertw_1}) the functional relations
satisfied by the kernel of the intertwining operator, written in integral
form and, consequently, the explicit representation of the matrix elements
of the operator itself.

It is known that the most degenerate principal series representations of $%
SO(N,1)$ labelled with the quantum number $j=-\frac{N-1}{2}+i\rho ,\ $ with $%
\rho >0,$\ can be realized on $\mathcal{L}_{2}\left( S^{N-1}\right) $ (see
Section 9.2.1 of Ref\cite{VK93}) 
\begin{equation}
U_{j}\left( g\right) f\left( \eta \right) =\left( \omega _{g}\right)
^{j}f\left( \eta _{g}\right) \;,\quad \eta \in S^{N-1}  \label{U_g1}
\end{equation}
where 
\[
\omega _{g}=\sum_{i=1}^{N}g_{Ni}^{-1}\eta _{i}+g_{NN},\quad \left( \eta
_{g}\right) _{k}=\frac{\sum_{i=1}^{N}g_{ki}^{-1}\eta _{i}+g_{kN}}{%
\sum_{i=1}^{N}g_{Ni}^{-1}\eta _{i}+g_{NN}} 
\]
The representations specified by labels $j\,\ $and $1-N-j\,$\ are
Weyl-equivalent.

The operator $A$ defined by 
\begin{equation}
\left( Af\right) \left( \eta \right) =\int \mathcal{K}\left( \eta ,\eta
^{\prime }\right) f\left( \eta ^{\prime }\right) d\eta ^{\prime }  \label{Af}
\end{equation}
intertwines representations $j$ and $1-N-j$ on condition that 
\begin{equation}
\mathcal{K}\left( \eta _{g},\eta _{g}^{\prime }\right) =\left( \omega
_{g}\right) ^{N-1+j}\left( \omega _{g}^{\prime }\right) ^{N-1+j}\mathcal{K}%
(\eta ,\eta ^{\prime })\;.  \label{funct}
\end{equation}

The kernel, $K$ , is uniquely determined by Eq. (\ref{funct}) up to a
constant and is given by 
\begin{equation}
\mathcal{K}(\eta ,\eta ^{\prime })=\varkappa \left( 1-\eta \cdot \eta
^{\prime }\right) ^{1-N-j}\;.  \label{kern}
\end{equation}
with 
\begin{equation}
\varkappa =2^{-\frac{N-1}{2}+{\normalsize i}\rho }\frac{\Gamma \left( \frac{%
N-1}{2}+{\normalsize i}\rho \right) }{\pi ^{\frac{N-1}{2}}\Gamma \left( -%
{\normalsize i}\rho \right) }
\end{equation}
Taking into account the fact that the spherical harmonics $Y_{lMK}$ (\ref%
{Harm}) form a basis in $\mathcal{L}_{2}\left( S^{3n-1}\right) $,
corresponding to the above reduction, we obtain the following integral
representation for the matrix elements of $A$%
\begin{equation}
\left\langle j;l^{\prime }M^{\prime }K^{\prime }\right| A\left|
j;lMK\right\rangle =\int \mathcal{K}(\eta ,\eta ^{\prime })Y_{l^{\prime
}M^{\prime }K^{\prime }}^{\ast }\left( \eta ^{\prime }\right) Y_{lMK}\left(
\eta \right) d\eta d\eta ^{\prime }\;.  \label{S_lK}
\end{equation}
Therefore 
\begin{equation}
\left\langle j;l^{\prime }M^{\prime }K^{\prime }\right| A\left|
j;lMK\right\rangle =A_{l}\delta _{ll^{\prime }}\delta _{MM^{\prime }}\delta
_{KK^{\prime }}\;,  \label{A_mat_el}
\end{equation}
where 
\begin{equation}
\;A_{l}=\frac{\Gamma \left( \frac{3n-1}{2}+{\normalsize i}\rho +l\right) }{%
\Gamma \left( \frac{3n-1}{2}-{\normalsize i}\rho +l\right) }.  \label{A_l}
\end{equation}

According to this, we have 
\begin{equation}
S\left( p;p^{\prime }\right) =\sum\limits_{lM}A_{l}\mathcal{Y}_{lM}\left( 
\hat{p}\right) \mathcal{Y}_{lM}^{\ast }\left( \hat{p}^{\prime }\right) \;.
\label{S_thetaphi}
\end{equation}

Thus, the scattering amplitude, $f\left( p;p^{\prime }\right) $, is defined
by 
\begin{equation}
f\left( p;p^{\prime }\right) =\left( -i\right) \left( \frac{2\pi }{p}\right)
^{\frac{n-1}{2}}\sum\limits_{lM}\left( A_{l}-1\right) \mathcal{Y}_{lM}\left( 
\hat{p}\right) \mathcal{Y}_{lM}^{\ast }\left( \hat{p}^{\prime }\right) \;.
\label{sc_ampl}
\end{equation}

We can omit unity in the brackets of formula (\ref{sc_ampl}) when $\hat{p}%
^{\prime }\neq \hat{p}$ , leaving 
\begin{equation}
f\left( p;p^{\prime }\right) =\left( -i\right) \left( \frac{2\pi }{p}\right)
^{\frac{n-1}{2}}\sum\limits_{lM}\frac{\Gamma \left( \frac{3n-1}{2}+%
{\normalsize i}\rho +l\right) }{\Gamma \left( \frac{3n-1}{2}-{\normalsize i}%
\rho +l\right) }\mathcal{Y}_{lM}\left( \hat{p}\right) \mathcal{Y}_{lM}^{\ast
}\left( \hat{p}^{\prime }\right) \;.  \label{sc_ampl_1}
\end{equation}

Moreover, formulae (\ref{Harm}) and the following expansion of the kernel 
\[
\left( 1-\eta \cdot \eta ^{\prime }\right) ^{-\frac{3n-1}{2}-{\normalsize i}%
\rho }=\left( 2\pi \right) ^{\frac{3n-1}{2}}\frac{2^{-i\rho }\Gamma \left( -%
{\normalsize i}\rho \right) }{\Gamma \left( \frac{3n-1}{2}+{\normalsize i}%
\rho \right) }\sum\limits_{lMK}\frac{\Gamma \left( \frac{3n-1}{2}+%
{\normalsize i}\rho +l\right) }{\Gamma \left( \frac{3n-1}{2}-{\normalsize i}%
\rho +l\right) }Y_{lMK}^{\ast }\left( \eta ^{\prime }\right) Y_{lMK}\left(
\eta \right) \;, 
\]
yield an integral representation of the scattering amplitude 
\begin{eqnarray}
f\left( p;p^{\prime }\right) &=&\frac{1}{ip^{\frac{n-1}{2}}}\frac{2^{i\rho
}\Gamma \left( \frac{3n-1}{2}+{\normalsize i}\rho \right) }{\Gamma \left( -%
{\normalsize i}\rho \right) }\Lambda \left( \theta _{n-1},\ldots ,\theta
_{1}\right) \Lambda \left( \theta _{n-1}^{\prime },\ldots ,\theta
_{1}^{\prime }\right)  \label{A_i} \\
&&\times \int_{0}^{\pi }\cdots \int_{0}^{\pi }\left( 1-\hat{p}_{1}\hat{p}%
_{1}^{\prime }\cos \alpha _{1}-\hat{p}_{2}\hat{p}_{2}^{\prime }\cos \alpha
_{2}-\cdots -\hat{p}_{n}\hat{p}_{n}^{\prime }\cos \alpha _{n}\right) ^{-%
\frac{3n-1}{2}-{\normalsize i}\rho }  \nonumber \\
&&\times P_{\kappa _{1}}\left( \cos \alpha _{1}\right) P_{\kappa _{2}}\left(
\cos \alpha _{2}\right) \cdots P_{\kappa _{n}}\left( \cos \alpha _{n}\right)
\sin \alpha _{1}\sin \alpha _{2}\cdots \sin \alpha _{n}d\alpha _{1}d\alpha
_{2}\cdots d\alpha _{n}  \nonumber
\end{eqnarray}

where 
\[
\Lambda \left( \theta _{n-1},\ldots ,\theta _{1}\right)
=\prod_{i=1}^{n-1}\sin ^{n-i}\theta _{n-i}\cos \theta _{n-i} 
\]
When $\kappa _{i}=0,$\ formula (\ref{A_i}) simplifies to 
\begin{equation}
f\left( p;p^{\prime }\right) =\frac{1}{ip^{\frac{n-1}{2}}}\frac{2^{i\rho
}\Gamma \left( \frac{n-1}{2}+{\normalsize i}\rho \right) }{\Gamma \left( -%
{\normalsize i}\rho \right) }\sum_{\epsilon _{i}=\pm }\sigma \left(
1-\epsilon _{1}\hat{p}_{1}\hat{p}_{1}^{\prime }-\epsilon _{2}\hat{p}_{2}\hat{%
p}_{2}^{\prime }-\cdots -\epsilon _{n}\hat{p}_{n}\hat{p}_{n}^{\prime
}\right) ^{-\frac{n-1}{2}-{\normalsize i}\rho }  \label{ampl_f}
\end{equation}
with 
\[
\sigma =\prod_{i=1}^{n}\epsilon _{i} 
\]

It is worth noting that the amplitude (\ref{ampl_f}) does not reduce to the
Coulomb amplitude $f_{Coul}$ in $n$ dimensions \cite{RS79} 
\begin{equation}
f_{Coul}\left( p;p^{\prime }\right) =\frac{1}{ip^{\frac{n-1}{2}}}\frac{%
2^{i\rho }\Gamma \left( \frac{n-1}{2}+{\normalsize i}\rho \right) }{\Gamma
\left( -{\normalsize i}\rho \right) }\left( 1-\hat{p}_{1}\hat{p}_{1}^{\prime
}-\hat{p}_{2}\hat{p}_{2}^{\prime }-\cdots -\hat{p}_{n}\hat{p}_{n}^{\prime
}\right) ^{-\frac{n-1}{2}-{\normalsize i}\rho }  \label{ampl_C}
\end{equation}%
when $\kappa _{i}$\ is set equal to zero. The reason for this discrepancy
lies in the fact that we solve the Schr\"{o}dinger equation for $x_{i}\geq
0,\ (i=1,2,\ldots ,n)$ \ with boundary conditions 
\[
\Psi \left( x\right) =0\textstyle{\ \ for\ \ }x_{i}=0,\;(i=1,\ldots ,n) 
\]

\section{Conclusions and outlook}

The present work is the latest in a series of papers where the potential
group approach and the method of intertwining operators (formulae (\ref%
{inv_group}-\ref{S_A_2})) have been applied to non-central extensions of the
Coulomb potential: Ref.\cite{Ke06} studied the bound states of the
three-dimensional Coulomb potential plus a barrier term with the $SO\left(
5\right) $ potential group and Ref.\cite{Ke07} the scattering states of
similar potentials with $SO\left( 5,1\right) $ potential group. A first
simultaneous analysis of bound and scattering states of a three-dimensional
Coulomb-Rosochatius potential of type (\ref{non_centr_V2}) was performed in
Ref\cite{KV10}, where the use of potential groups $SO\left( 7\right) $ for
bound states and $SO\left( 6,1\right) $ for scattering states. Generally, in
the $n$-dimensional case for Coulomb-Rosochatius potential one can choose
potential groups $SO\left( qn+1\right) $ and $SO\left( qn,1\right) $ with $%
q=2,3,\ldots $\ . Then the potential strengths $\beta _{i}$ will be related
to eigenvalues $k_{i}\left( k_{i}+q-2\right) ,\ k_{i}=0,1,2,\ldots ,$ of
Casimir operators of subgroups $SO\left( q\right) _{i}$ as: $\beta
_{i}=\left( k_{i}+\frac{q-2}{2}\right) ^{2}-\frac{1}{4}.$\ In the present
work, we have chosen $q=3$.

The method has proved useful also in deriving the full set of \ $2n-1$\
constants of motion, related to Casimir invariants of subgroups appearing in
the decomposition chains, thus proving the maximal superintegrability of the
system.

The potential group approach is quite general and, obviously, not limited to
the orthogonal and pseudo-orthogonal symmetries underlying the
Coulomb-Rosochatius Hamiltonian. Systems with different symmetries will be
studied in future works.

%\section*{References}

\end{document}